\documentclass[aps,prb,showpacs,reprint,superscriptaddress]{revtex4-2}
\usepackage{graphicx}
\usepackage{color}
\usepackage{amsmath}
\usepackage{bbm}
\usepackage{amssymb}
\usepackage{hyperref}
\usepackage{placeins}  
\usepackage{braket} 
\usepackage{float}
\DeclareFontFamily{OT1}{pzc}{}
\DeclareFontShape{OT1}{pzc}{m}{it}{<-> s * [1.10] pzcmi7t}{}
\DeclareMathAlphabet{\mathpzc}{OT1}{pzc}{m}{it}

\usepackage{dsfont}

\usepackage[utf8]{inputenc}	% kodowanie UTF-8
\usepackage[T1]{fontenc}
\usepackage{bm}

\usepackage[svgnames]{xcolor}

\input epsf

\begin{document}

\title{Interplay between topology and electron-electron interactions in the moir\'{e} MoT\texorpdfstring{e\textsubscript{2}}{e2}/WS\texorpdfstring{e\textsubscript{2}}{e2} heterobilayer}

% Force line breaks with \\

\author{Palash Saha}
\affiliation{Academic Centre for Materials and Nanotechnology, AGH University of Krakow, Al. Mickiewicza 30, 30-059 Krakow, Poland}%
\author{Louk Rademaker}
\affiliation{Department of Quantum Matter Physics, University of Geneva, CH-1211 Geneva, Switzerland}
\author{Micha{\l} Zegrodnik}%
 \email{michal.zegrodnik@agh.edu.pl}
\affiliation{Academic Centre for Materials and Nanotechnology, AGH University of Krakow, Al. Mickiewicza 30, 30-059 Krakow, Poland}%

\date{\today}

\begin{abstract}
We study, the interplay between topology and electron-electron interactions in the moir\'{e}  MoTe\(_2\)/WSe\(_2\) heterobilayer. In our analysis we apply an effective two-band model with complex hoppings that incorporates the Ising-type spin-orbit coupling and lead to a non-trivial topology after the application of perpendicular electric field (displacement field). The model is supplemented by on-site and inter-site Coulomb repulsion terms and treated by both Hartree-Fock and Gutzwiller methods. According to our analysis, for the case of one hole per moir\'{e} unit cell, the system undergoes two phase transitions with increasing displacement field. The first one is from an in-plane 120$^\circ$ antiferromagnetic charge transfer insulator to a topological insulator. At the second transition, the system becomes topologically trivial and an out-of-plane ferrimagnetic metallic phase becomes stable. In the topological region a spontaneous spin-polarization appears and the holes are distributed in both layers. Additionally, we analyze the influence of the intersite Coulomb repulsion terms on the appearance of the topological phase as well as on the formation of the charge density wave state. We discuss the obtained results in the context of available experimental data.
\end{abstract}

\maketitle

\section{\label{sec:level1}Introduction}

In recent years, the moir\'{e} transition metal dichalcogenide (TMD) bilayers have emerged as a promising platform to study the interplay between non-trivial topology, strong electron-electron correlations, and spin-orbit coupling. In those systems the moir\'{e} pattern is obtained due to rotational misalignment, as in the twisted WSe$_2$ bilayer, or due to lattice mismatch, as in the MoTe$_2$/WSe$_2$ heterobilayer. For the latter case and at zero out-of-plane electric field, the system is a Mott (or charge-transfer) insulator with one hole per moir\'{e} unit cell\cite{Ghiotto2021,Wang2020,Zhao2023}. This effect points to a significant role played by the electron-electron interactions which are relatively large in comparison with the single-particle energy, due to the appearance of flat electronic bands. Interestingly, within some range of non-zero perpendicular electric fields a quantum anomalous Hall insulator (QAHI) is developed\cite{mw_exp_1,mw_exp_2,Tao2024} which indicates non-trivial topology. By probing the magnetic properties, it has been determined that the QAHI ground state is characterized by a spontaneous spin polarization and holes distributed in both TMD layers\cite{mw_exp_2}. 

The physics of the MoTe$_2$/WSe$_2$ heterobilayer is determined by the two moir\'{e} valence bands, with the first band originating from a Wannier orbital centered at the $MM$ stacking in the MoTe$_2$ layer, and the second band originating from an orbital at the $XX$ stacking point in the WSe$_2$ layer. Here $M$=Mo,W and $X$=Te,Se. This results in a two-sublattice honeycomb structure with the $MM$ and $XX$ orbitals sitting at the two lattice sites of a single unit cell. Along these lines, a two-band model has been formulated with complex intra- and inter-layer hoppings which incorporate the Ising-type spin-valley locking in each of the two bands\cite{louk,mw2}. Additionally, by changing the out-of-plane electric field (displacement field) one can tune the relative energy of the two bands. In particular, for large enough values of the displacement field, band inversion appears leading to non-trivial topology. For two holes per moir\'{e} unit cell, when the upper band is empty, such physical picture leads to transition from a band insulator to a quantum spin Hall insulator (QSH) with increasing displacement field, which is in agreement with the experimental reports\cite{r2}. 

The observed transition from the Mott (or charge transfer) insulator to QAHI at one hole per moir\'{e} unit cell requires going beyond single-particle physics. In such case one has to supplement the model at least with the onsite Coulomb interaction terms. This leads to a Hubbard model description, in which both correlation effects and non-trivial topology play a significant role. Such description has been theoretically investigated in the context of interplay between the magnetically ordered states and topological properties\cite{mw2} as well as Kondo physics\cite{mw1,Xi_2024,Millis_Kondo_2024,Chowdhury_2024}. Also, both the Hubbard and $t$-$J$ models have been recently applied to propose that an excitonic Chern insulator can be realized in MoTe$_2$/WSe$_2$\cite{r3} {and interlayer excitonic physics has been discussed in the context of possible pairing mechanism in Ref. \cite{Millis2023_SC_heterobilayer}.} Theoretical investigations of various magnetically and charge ordered states have also been carried out with the use of a plane-wave method without the projection to a few selected moir\'{e} bands\cite{mw3,Chang2022}.

In this paper we study the extended Hubbard Hamiltonian appropriate for an effective description of the AB-stacked MoTe$_2$/WSe$_2$ heterobilayer and focus on the evolution of the electronic properties of the system at half-filling with increasing displacement field. We aim at reproducing the recent experimental data which indicate two critical values of the displacement field: the first one corresponds to a transition from a Mott/charge transfer insulating to a Chern insulating state, while at the second critical value the system becomes topologically trivial showing a metallic behavior. We also analyze the hole distribution across the layers together with magnetic ordering, and discuss the obtained results with the available experimental data. Additionally, we supplement our paper with the analysis of the charge-density-wave formation induced by the long-range Coulomb repulsion term away from half-filling. The bulk of the calculations are carried out within the mean-field Hartree-Fock method. However, since we reach relatively large values of Coulomb repulsion terms we also provide corresponding results stemming from the Gutzwiller approximation method.

Our main result is summarized in Fig.~\ref{mean_n}. At a density of one hole per moir\'{e} unit, we first find an insulating state with 120$^\circ$ antiferromagnetic order, with the spin moments completely localized in the MoTe$_2$ layer. Increasing the displacement field induces a transition to a quantum anomalous Hall state whereby the spin become canted in the $z$ direction. This is consistent with the picture obtained from recent experiments\cite{Tao2024}. When the system loses its antiferromagnetic order, a second transition is induced towards a ferrimagnetic metal.

The paper is organized as follows. In Sec. II we provide the details of the model itself and show the implementation of the the Hartree-Fock and Gutzwiller approaches. In Sec. IIIA we focus on the analysis of the topological and magnetic features of the system limiting to the onsite Coulomb repulsion only, whereas, the influence of the intersite repulsion as well the analysis of the charge ordered states is deferred to Sec. IIIB. Finally, in Section IV we summarize and conclude our most important results. Additionally, the details of the Chern number calculations as well as the obtained Berry curvature maps are presented in the Appendix A while the influence of the interlayer coherence terms is analyzed in Appendix B.
%%%%%%%%%%%%%%%%%%%%%%%%%%%%%%%%%%%%%%%%%%%%%%%%%%%%%%%%%%%%%%%%%%%%%%%%%%%%%%%%%%%%%%%%%%%%%%%%%%%%%%%%%%%%%%%%
\vspace{0.5cm}
\section{Model And Method}
\begin{figure}
\includegraphics[width=0.95\linewidth, height=3.75cm]{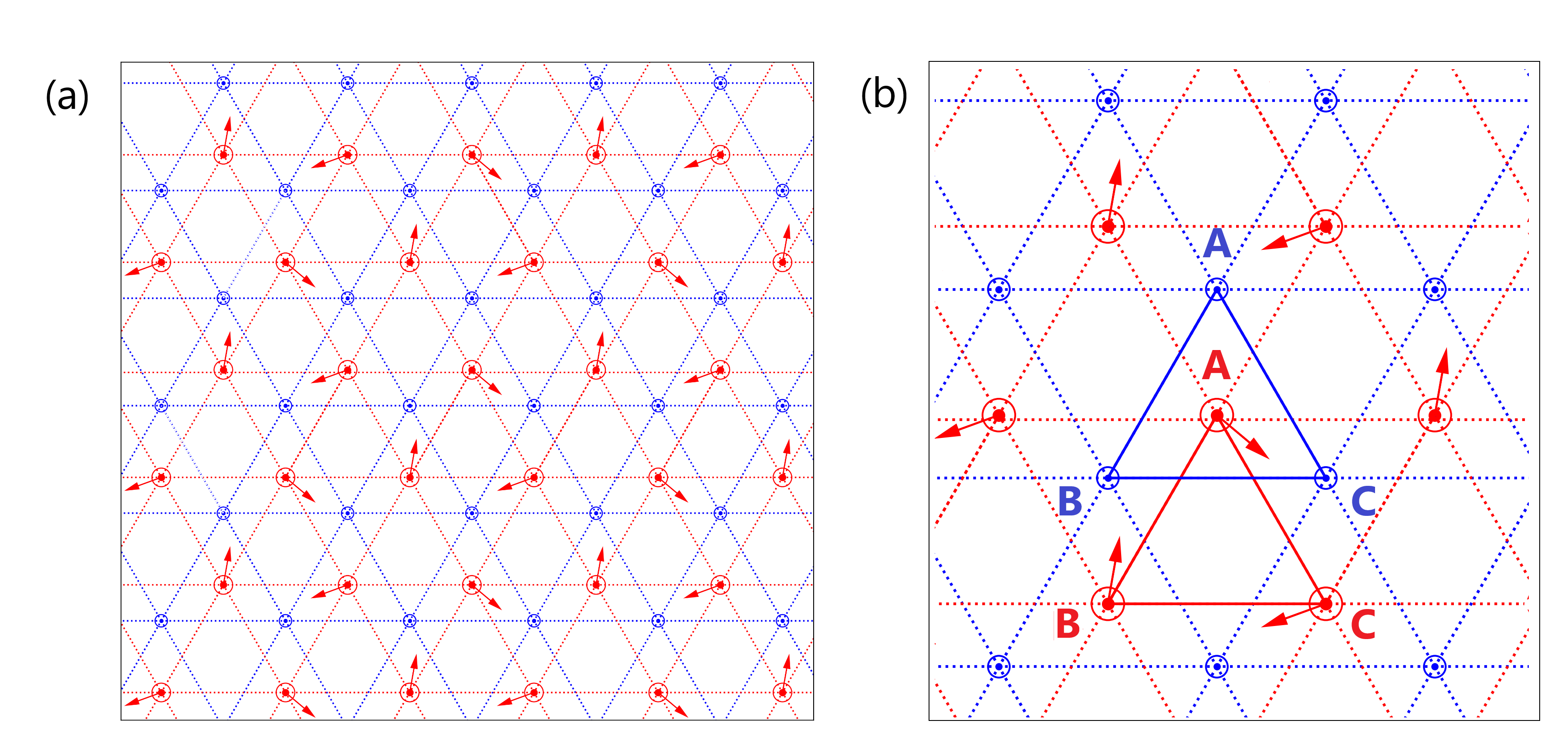}
\caption{(a) The two-sublattices of the resulting honeycomb lattice are marked in red (blue) and correspond to the MoTe$_2$ (WSe$_2$) layers. The schematic representation of the $120^{\circ}$ degree AF alignment at the MoTe$_2$ layer is provided by arrows.  (b) In order to take into account the possible $120^{\circ}$ degree AF alignment in both layers we consider a supercell containing 3 lattice sites per each layer (marked by $A,B$, and $C$).}
\label{af_lattice}
\end{figure}

We start with the two-band extended Hubbard Hamiltonian of the following form
 %H_tUV
 \begin{eqnarray}
 \mathcal{H}_{tU\mathcal{V}}=\mathcal{H}_{t} +\mathcal{H}_{U}+\mathcal{H}_{\mathcal{V}},
\label{ham0}
\end{eqnarray}
where the first term represents the tight-binding model provided in Ref. \onlinecite{louk}, which describes effectively the bare band structure of the AB-stacked MoTe$_2$/WSe${_2}$ heterobilayer, 
{\begin{equation}
\begin{split}
\hat{\mathcal{H}}_{t} &= \sum_{\langle\langle ij\rangle\rangle l\sigma} t^{l}_{ij\sigma}\;\hat{c}_{i l \sigma}^{\dagger}\; \hat{c}_{j l' \sigma}^{} + \sum_{\langle ij\rangle \sigma} \big(t^{12}_{ij\sigma\bar{\sigma}}\;\hat{c}_{i 1 \sigma}^{\dagger}\; \hat{c}_{j 2 \bar{\sigma}}^{}+H.c.\big)\\
&+(V+\Delta)\sum_{i}\hat{n}_{il=2},
\end{split}
\label{ham1}
\end{equation} }
where $\hat{c}_{i l \sigma}^{\dag}$ ($\hat{c}_{i l \sigma}$) creates (annihilates) an electron at the Wannier orbital centered at the $i$-th lattice site of sublattice $l$ with spin $\sigma$. Index $l$ distinguishes between two triangular sublattices consisting of the $MM$ ($l=1$) and $XX$ ($l=2$) stacking points in the MoTe$_2$ and WSe$_2$ layers, respectively (with $M=Mo,W$ and $X=Te,Se$). This results in a two-sublattice honeycomb structure [cf. Fig. \ref{af_lattice}(a)]. 
The $\langle\langle ij\rangle\rangle$ summation corresponds to next-nearest-neighbor (intralayer) hopping at the honeycomb lattice, while $\langle ij\rangle$ represents nearest-neighbor hopping, connecting lattice sites from different layers only. The intra- and inter-layer hoppings are complex and have the following form
\begin{equation}
    t^{l}_{ij\sigma}=t_l\; e^{i \phi \sigma^z \nu_{ij}} 
\end{equation}
\begin{equation}
    t^{12}_{ij\sigma\bar{\sigma}}=t_{\perp}\; e^{-i \phi \sigma^z \eta_{ij}}
\end{equation}
where $\phi=2\pi/3$ with $\sigma^z=1$ ($\sigma^z=-1$) for $\sigma=\uparrow$ ($\sigma=\downarrow$). Moreover, $\nu=\pm1$ depending on the direction of the hopping while $\eta=0,1,2$ and increases counterclockwise when going around the MoTe$_2$ lattice sites. The absolute values of the hopping amplitudes are set to $|t_1|$=4.03 meV, $|t_2|$=3.4 meV and $|t_{\perp}|$=4 meV and have been taken from Ref.~\onlinecite{louk}. 
{ We note that other studies \cite{mw2,mw5,Xi_2024} consider slightly different hopping amplitudes, e.g. $t_2\approx 9$ meV , $t_1 \approx 4.5$ meV, and $t_{\perp}\approx1.5-2$ meV are used in Ref. \onlinecite{mw5}. The differences likely arise due to methodological variations in the tight-binding fitting and/or the details of the DFT calculations. However, the hoppings taken in Ref. \onlinecite{mw5} and those used here are of the same order, therefore should not lead to qualitative changes in the resulting physical picture.}
\begin{figure}[h]
\includegraphics[width=0.9\linewidth, height=4.5 cm]{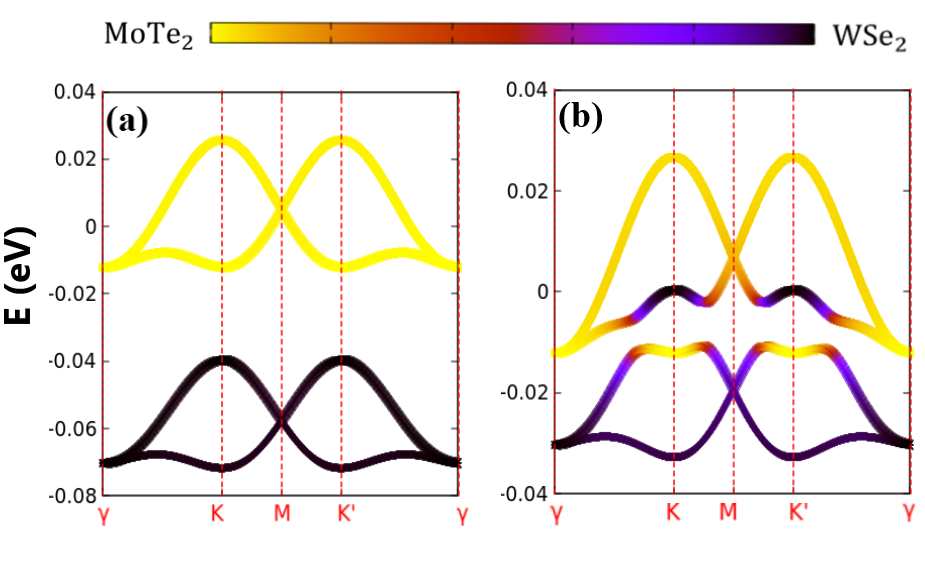}
\caption{The bare band structure defined by Eq. (\ref{ham1}) for two selected values of the displacement field: (a) $V=0$ eV and (b) $V=0.04$ eV. The colored scale indicates the contribution of the MoTe$_2$ and WSe$_2$ states to the resulting bands. Note that by increasing the displacement field one pushes the bottom band into the upper one, initiating band inversion.}
\label{two_bands}
\end{figure}

{It is worth emphasizing that by nature of the AB-stacking this model includes spin-flip interlayer hopping, as the spin-up valley in MoTe$_2$ is aligned with the spin-down valley in WSe$_2$. In a lowest order continuum model without interactions or lattice relaxation, such spin-flip hopping would be vanishingly small. However, it has been argued that it can arise through lattice relaxation~\cite{r2} or interaction effects~\cite{r3}. In this work we take the existence of interlayer spin-flip hopping as a starting point to study the effect of further interactions, without discussing the origin of such hopping.}

The last term of Eq. (\ref{ham1}) represents the onsite contribution where $\hat{n}_{il}=\hat{n}_{il\uparrow}+\hat{n}_{il\downarrow}$, $\hat{n}_{il\sigma}=\hat{c}_{il\sigma}^{\dag}\hat{c}_{il\sigma}^{}$, $\Delta=-60$ meV, and $V$ introduces the effect of displacement field. For $V=0$ the top (bottom) band is composed only of the MoTe$_2$ (WSe$_2$) states while with increasing $V$ one pushes the bottom band to the upper which at some point initiates band inversion (cf. Fig. \ref{two_bands}).

{It should be noted that, in general, the unit cell of the considered mor{\'e} heterobilayer contains a large number of atoms, leading to a multiband electronic structure. However, for the sake of simplicity, such description can be reduced by projecting onto selected bands and creating an effective lattice model\cite{louk,Xi_2024}. In principle, within such a description, longer-range hopping terms above the second nearest-neighbor can appear. However, as shown in Ref. \cite{louk}, those additional hoppings are one order of magnitude smaller than those which have been included by us here. Therefore, we limit here to hopping terms up to the second nearest neighbor.}

The second and the third terms in Eq. (\ref{ham0}) introduce the onsite and intersite Coulomb repulsion and have the following form,
\begin{eqnarray}
\hat{\mathcal{H}}_{U} = U\sum_{il}{ \hat{n}_{il\uparrow} \hat{n}_{il\downarrow} },
\label{ham2}
\end{eqnarray}
\begin{eqnarray}
\hat{\mathcal{H}}_{\mathcal{V}} = \sum_{\langle\langle ij\rangle\rangle l \ }\mathcal{V}_{\parallel}{ \hat{n}_{il} \hat{n}_{jl}}+\sum_{\langle ij\rangle  \ }\mathcal{V}_{\perp}{ \hat{n}_{i1} \hat{n}_{j2}},
\label{ham3}
\end{eqnarray}
where $U$ determines the strength of the onsite repulsion while $\mathcal{V}_{\parallel}$ and $\mathcal{V}_{\perp}$ correspond to the intra- and inter-layer Coulomb repulsion, respectively.

We implement the Hartree-Fock (HF) approximation for the interaction terms. The principal insights derived from our HF investigation are firmly validated by subsequent variational Gutzwiller approach, which we detail further below. After applying the HF treatment to the onsite interaction term given by Eq. (\ref{ham2}), one obtains,
\begin{align}
\hat{\mathcal{H}}_{U} \approx & \, U\sum_{il}\big(\hat{n}_{il\uparrow}n_{il\downarrow} + \hat{n}_{il\downarrow}n_{il\uparrow} - \hat{S}_{il}^{+}S_{il}^{-} \nonumber\\
& - \hat{S}_{il}^{-}S_{il}^{+} - n_{il\uparrow}n_{il\downarrow} + S_{il}^+S_{il}^-\big),
\label{ham4}
\end{align}
where, $\hat{S}_{il}^{+}=\hat{c}_{il\uparrow}^{\dag}\hat{c}_{il\downarrow}$ and $\hat{S}_{il}^{-}=\hat{c}_{il\downarrow}^{\dag}\hat{c}_{il\uparrow}$. In the above equation we have introduced the following notation for the expectation values: $S_{il}^{\pm}=\langle\hat{S}_{il}^{\pm}\rangle$ and $n_{il\sigma}=\langle\hat{n}_{il\sigma}\rangle$. 

Application of the same approach to the intersite Coulomb repulsion given by Eq. (\ref{ham3}) leads to,
{
\begin{eqnarray}
\hat{\mathcal{H}}_{\mathcal{V}_{\parallel}} =\sum_{\substack{\langle\langle ij\rangle\rangle l\\ \sigma \sigma'  } }\mathcal{V}_{\parallel}\big({ \hat{n}_{il\sigma}n_{jl\sigma'} - \frac{1}{2}{n}_{il\sigma}n_{jl\sigma'}  }\big),
\label{ham_V_HF}
\end{eqnarray}
\begin{equation}
\begin{split}
\hat{\mathcal{H}}_{\mathcal{V}_{\perp}} &=\sum_{\substack{\langle ij\rangle l\\\sigma \sigma'  } }\mathcal{V}_{\perp}\big({ \hat{n}_{il\sigma}n_{j\bar{l}\sigma'} - \frac{1}{2}{n}_{il\sigma}n_{j\bar{l}\sigma'}  }\big)\\
&-\sum_{\substack{\langle ij\rangle l\\\sigma \sigma'  } }\mathcal{V}_{\perp}\big(\hat {P}_{i\sigma j\sigma'}^{l\bar{l}} {P}_{j\sigma' i\sigma}^{\bar{l}l} -\frac{1}{2}{P}_{i\sigma j\sigma'}^{l\bar{l}} {P}_{j\sigma' i\sigma}^{\bar{l}l}
\big),
\end{split}
\label{ham_V_HF2}
\end{equation}}
where $n_{il}=n_{il\uparrow}+n_{il\downarrow}$, { $\hat{{P}}_{i\sigma j\sigma'}^{l\bar{l}} = \hat{c}_{i l \sigma}^{\dagger} \hat{c}_{j \bar{l} \sigma'}^{}$ and ${P}_{i\sigma j\sigma'}^{l\bar{l}}=\langle\hat{c}_{i l \sigma}^{\dagger} \hat{c}_{j \bar{l} \sigma'}^{}\rangle$. 
We take into account both spin-conserving and spin-flip interlayer coherence terms in the HF decomposition of the $\sim\mathcal{V}_{\perp}$ term, to make sure that we determine the true ground state of the considered bilayer problem.}

%&-\sum_{\substack{\langle ijll'\rangle\\ \sigma  } }\mathcal{V}_{ll'}\big(\hat {P}_{iljl'\sigma} {P}_{jl'il\sigma} -\frac{1}{2}{P}_{iljl'\sigma} {P}_{jl'il\sigma}

As one can see from Eq. (\ref{ham4}), already at the level of such mean-field picture, the $U$-term can lead to effective shift of the spin-up and spin-down onsite energies as well as to non-zero expectation values of the onsite spin raising and/or lowering operators. That points to a possible spin-ordered states that can be stabilized due to electron-electron interaction effects. In this study, we focus on the situation corresponding to one hole per moir\'{e} unit cell. In such case and for $V=0$, one can project out the completely filled WSe$_2$ band, which is separated by an energy gap from the half-filled MoTe$_2$. That leads to an effective single-band model on a triangular lattice with nearest neighbor complex hoppings\cite{louk}. It has been reported previously that in such situation the onsite Coulomb repulsion leads to a tendency towards an in-plane 120$^\circ$ antiferromagnetic alignment\cite{Biborski2024}. Additionally, for sufficiently high value of $U$ the single-band model can be transformed into a Heisenberg model supplemented with the Dzyaloshinskii-Moriya (DM) term which also points to a possible antiferromagnetic ordering\cite{dm}. Based on those considerations, one should expect that with $V=0$, also for the two band model considered here, the in-plane 120$^\circ$ AF alignment should appear at the MoTe$_2$ layer [cf. Fig.\ref{af_lattice}(a)]. Such ordered state can than evolve with increasing value of the displacement field. To allow for an initial AF alignment in our calculations we introduce a supercell containing 3 moir\'{e} unit cells, each consisting of one lattice site from the MoTe$_2$ layer and one from the WSe$_2$ layer [bold lines in Fig.\ref{af_lattice}(b)]. Now, the expectation values of the occupation operator as well as the spin raising and lowering operators can vary within the supercell, but they need to repeat themselves when moving from one supercell to another. Namely, $n_{ilp\sigma}\equiv n_{lp\sigma}$, $n_{ilp}\equiv n_{lp}$, $S^{\pm}_{ilp}\equiv S^{\pm}_{lp}$, where $i$ enumerates the supercells, $p=A,B,C$ determines the lattice site within the supercell with $l=1,2$ correspond to the two layers. In such notation the number of electrons per MoTe$_2$ and WSe$_2$ moir\'{e} orbitals is: $n_1=(n_{1A}+n_{1B}+n_{1C})/3$ and $n_2=(n_{2A}+n_{2B}+n_{2C})/3$, respectively. Also, the total number of particles per lattice site takes the form $n_{tot}=n_1+n_2$. The relation between $n_{lp\sigma}$, $S^{\pm}_{lp}$ and the components of the magnetization vectors are the following
\begin{equation}
\begin{split}
    S_{lp}^x &= \frac{1}{2}(S^+_{lp}+S^-_{lp}),\\
    S_{lp}^y &= \frac{1}{2i}(S^+_{lp}-S^-_{lp}),\\
    S_{lp}^z &= \frac{1}{2}(n_{lp\uparrow}-n_{lp\downarrow}).\\
\end{split}
\end{equation}
In principle, such approach allows for different orientations of the magnetization vector, $\mathbf{S}_{lp}=(S^x_{lp},S^y_{lp},S^z_{lp})$, at each of the 6 lattice sites of the chosen supercell. In particular, the mentioned in-plane $120^{\circ}$ antiferromagnetic alignment at the MoTe$_2$ layer can be realized, for which all the three vectors $\mathbf{S}_{lp}$, for $p=A,B,C$ and $l=1$, have equal module, no $z$-component, and they form an angle of $120^{\circ}$ between each other. 

In order to be able to effectively diagonalize the mean-field hamiltonian composed of Eqs. (\ref{ham1}), (\ref{ham4}), (\ref{ham_V_HF}), and (\ref{ham_V_HF2}), one can first carry out the transformation to the reciprocal space, which leads to
\begin{equation}
\begin{split}
    \mathcal{H} &= \sum_{\mathbf{k}} \mathbf{f}_{\mathbf{k}}^{\dag}\mathbf{H}_{\mathbf{k}}^{} \mathbf{f}_{\mathbf{k}}^{} + U N \sum_{pl}\big(n_{pl\uparrow} n_{pl\downarrow} + S_{pl}^{+} S_{pl}^{-}\big) \\
    &\quad - \mathcal{V}_{\parallel}\frac{3N}{2}\sum_{\substack{pp'l \\ (p \neq p')}} n_{pl} n_{p'l} - \mathcal{V}_{\perp} N \sum_{pp'} n_{p1} n_{p'2}\\
    &\quad+ \mathcal{V}_{\perp} \frac{N}{2} \sum_{\substack{pp' \\ \sigma \sigma'}} P_{p\sigma p'\sigma'}^{12} P_{p'\sigma'p\sigma}^{21},    
\end{split}
\label{ham6}
\end{equation}
where $N$ is the number of supercells in the system and we have kept the scalar terms in the real-space representation. Also, we introduced a $12$-dimensional composite creation and annihilation operators in $\mathbf{k}$-space defined by
\begin{align}
\mathbf{f}^{\dag}_{\mathbf{k}} = & \
(\hat{c}_{A\mathbf{k}1\uparrow}^{\dag} \, \hat{c}_{B\mathbf{k}1\uparrow}^{\dag} \, \hat{c}_{C\mathbf{k}1\uparrow}^{\dag} \, \hat{c}_{A\mathbf{k}2\uparrow}^{\dag} \, \hat{c}_{B\mathbf{k}2\uparrow}^{\dag} \, \hat{c}_{C\mathbf{k}2\uparrow}^{\dag} \nonumber \\
& \, \hat{c}_{A\mathbf{k}1\downarrow}^{\dag} \, \hat{c}_{B\mathbf{k}1\downarrow}^{\dag} \, \hat{c}_{C\mathbf{k}1\downarrow}^{\dag} \, \hat{c}_{A\mathbf{k}2\downarrow}^{\dag} \, \hat{c}_{B\mathbf{k}2\downarrow}^{\dag} \, \hat{c}_{C\mathbf{k}2\downarrow}^{\dag}),
\end{align}
and $
\mathbf{f}_{\mathbf{k}} = (\mathbf{f}^{\dagger}_{\mathbf{k}})^{\dagger}
$. Note that, the three moir\'{e} lattice sites per each layer are labeled by \(A, B,\) and \(C\), with distinctions between layers indicated by the respective layer indices, $l=1,2$ [cf. Fig. \ref{af_lattice} (b)].
{
The general matrix form of the Hamiltonian appearing in Eq. (\ref{ham6}) is the following
%H_matrix
\renewcommand{\arraystretch}{1.5}
\begin{eqnarray}
\mathbf{H}_\mathbf{k}=
\left(
\begin{array}{cc}
\mathbf{H}_{\uparrow}(\mathbf{k})   &  \mathbf{H}_{\uparrow\downarrow}(\mathbf{k})\\ 
   \mathbf{H}_{\downarrow\uparrow}(\mathbf{k})  & \mathbf{H}_{\downarrow}(\mathbf{k})
\end{array}
\right),
\label{ham7}
\end{eqnarray}
where 
\begin{equation}
 \mathbf{H}_{\uparrow\downarrow}(\mathbf{k})=U\mathbf{S}^{-}-\mathcal{V}_{\perp}\mathbf{P}^{-} +\mathbf{T}_{\uparrow\downarrow}(\mathbf{k}),
 \label{eq:spin_mixing_martix}
\end{equation}
and $\mathbf{H}_{\downarrow\uparrow}(\mathbf{k})=(\mathbf{H}_{\uparrow\downarrow}(\mathbf{k}))^*$. The $\mathbf{S}^{-}$ matrices  correspond to the onsite spin-flip operators and result from the onsite Coulomb repulsion,
\begin{equation}
\begin{aligned}
\mathbf{S}^{-}=
\left(
\begin{array}{c}
\mathbf{S}^{-}_1\oplus\mathbf{S}^{-}_2
\end{array}
\right);
\mathbf{S}^{-}_{l}=
\left(
\begin{array}{ccc}
S^{-}_{Al}(\mathbf{k})   & 0 & 0 \\
   0  &S^{-}_{Bl}(\mathbf{k}) & 0 \\
      0  & 0 & S^{-}_{Cl} (\mathbf{k})
\end{array}
\right).
\label{ham60}
\end{aligned}
\end{equation}
} 
{Above terms are responsible for the tendency towards in-plane 120$^{\circ}$ AF alignment. The $\mathbf{P}^{-}$ matrix originates from the interlayer Coulomb repulsion and have the form,
\begin{equation}
\begin{aligned}
&\hspace{3em}\mathbf{P}^{-} = 
\left(
\begin{array}{cc}
0 & \mathbf{P^{-}_{21}} \\
\mathbf{P^{-}_{12}} & 0
\end{array}
\right);\\[2ex]
\mathbf{P}^{-}_{l\bar{l}} &= 
\scalebox{1}{$
\left(
\begin{array}{ccc}
P^{l\bar{l}}_{A\downarrow A\uparrow}(\mathbf{k}) & P^{l\bar{l}}_{B\downarrow A\uparrow}(\mathbf{k}) & P^{l\bar{l}}_{C\downarrow A\uparrow}(\mathbf{k}) \\
P^{l\bar{l}}_{A\downarrow B\uparrow}(\mathbf{k}) & P^{l\bar{l}}_{B\downarrow B\uparrow}(\mathbf{k}) & P^{l\bar{l}}_{C\downarrow B\uparrow}(\mathbf{k}) \\
P^{l\bar{l}}_{A\downarrow C\uparrow}(\mathbf{k}) & P^{l\bar{l}}_{B\downarrow C\uparrow}(\mathbf{k}) & P^{l\bar{l}}_{C\downarrow C\uparrow}(\mathbf{k}) 
\end{array}
\right)$},
\end{aligned}
\end{equation}
 where ${{P}}_{p\sigma p'\sigma'}^{l\bar{l}}(\mathbf{k}) = \langle{c}_{p\mathbf{k} l\sigma}^{\dagger} {c}_{p' \mathbf{k}\bar{l}\sigma'}^{}\rangle$. The last term in Eq. (\ref{eq:spin_mixing_martix}) originates from the single particle part of the Hamiltonian and is the following}
\begin{equation}
\begin{aligned}
&\hspace{3em}\mathbf{T}_{\uparrow\downarrow} = 
\left(
\begin{array}{cc}
\mathbf{T^{12}_{\uparrow\downarrow}} & 0 \\
0 & \mathbf{T^{21}_{\uparrow\downarrow}}
\end{array}
\right);\\[2ex]
\mathbf{T}^{l\bar{l}}_{\uparrow\downarrow} &= 
\scalebox{1}{$
\left(
\begin{array}{ccc}
t^{l\bar{l}}_{A\uparrow A\downarrow}(\mathbf{k}) & t^{l\bar{l}}_{B\uparrow A\downarrow}(\mathbf{k}) & t^{l\bar{l}}_{C\uparrow A\downarrow}(\mathbf{k}) \\
t^{l\bar{l}}_{A\uparrow B\downarrow}(\mathbf{k}) & t^{l\bar{l}}_{B\uparrow B\downarrow}(\mathbf{k}) & t^{l\bar{l}}_{C\uparrow B\downarrow}(\mathbf{k}) \\
t^{l\bar{l}}_{A\uparrow C\downarrow}(\mathbf{k}) & t^{l\bar{l}}_{B\uparrow C\downarrow}(\mathbf{k}) & t^{l\bar{l}}_{C\uparrow C\downarrow}(\mathbf{k}) 
\end{array}
\right)$},
\end{aligned}
\end{equation}
where the $t_{p\sigma p'\sigma'}^{l\bar{l}}$ correspond to the standard momentum space representation of the single particle part of our starting Hamiltonian corresponding to the inter-layer spin-mixing hopping contributions between the $A$, $B$, and $C$ lattice sites [ cf. Eq. (\ref{ham1})].
 
Finally, each spin-conserving submatrices in Eq. (\ref{ham7}) are defined as,
%H_spin_matrix
\begin{equation}
\mathbf{H}_{\sigma}(\mathbf{k})=\mathbf{T}_{\sigma}(\mathbf{k})+ (\mathbf{I}_{1\sigma}\oplus\mathbf{I}_{2\sigma}),
\end{equation}
\vspace{0.5cm}
where $\mathbf{T}_{\sigma}(\mathbf{k})$ takes the form {
\begin{flushleft}
$\mathbf{T}_{\sigma}(\mathbf{k}) =$
\end{flushleft}
\begin{equation}
\renewcommand{\arraystretch}{1.5}
\begin{aligned}
 %\\[1ex]
{\small
\left(
\begin{array}{cccccc}
0 & \mathit{t}^{11}_{AB\sigma}(\mathbf{k}) & \mathit{t}^{11}_{AC\sigma}(\mathbf{k}) & \mathfrak{\tau}^{12}_{AA\sigma}(\mathbf{k}) & \mathfrak{\tau}^{12}_{AB\sigma}(\mathbf{k}) & \mathfrak{\tau}^{12}_{AC\sigma}(\mathbf{k}) \\
\mathit{t}^{11}_{BA\sigma}(\mathbf{k}) & 0 & \mathit{t}^{11}_{BC\sigma}(\mathbf{k}) & \mathfrak{\tau}^{12}_{BA\sigma}(\mathbf{k}) & \mathfrak{\tau}^{12}_{BB\sigma}(\mathbf{k}) & \mathfrak{\tau}^{12}_{BC\sigma}(\mathbf{k}) \\
\mathit{t}^{11}_{CA\sigma}(\mathbf{k}) & \mathit{t}^{11}_{CB\sigma}(\mathbf{k}) & 0 & \mathfrak{\tau}^{12}_{CA\sigma}(\mathbf{k}) & \mathfrak{\tau}^{12}_{CB\sigma}(\mathbf{k}) & \mathfrak{\tau}^{12}_{CC\sigma}(\mathbf{k}) \\
\mathfrak{\tau}^{21}_{AA\sigma}(\mathbf{k}) & \mathfrak{\tau}^{21}_{AB\sigma}(\mathbf{k}) & \mathfrak{\tau}^{21}_{AC\sigma}(\mathbf{k}) & 0 & \mathit{t}^{22}_{AB\sigma}(\mathbf{k}) & \mathit{t}^{22}_{AC\sigma}(\mathbf{k}) \\
\mathfrak{\tau}^{21}_{BA\sigma}(\mathbf{k}) & \mathfrak{\tau}^{21}_{BB\sigma}(\mathbf{k}) & \mathfrak{\tau}^{21}_{BC\sigma}(\mathbf{k}) & \mathit{t}^{22}_{BA\sigma}(\mathbf{k}) & 0 & \mathit{t}^{22}_{BC\sigma}(\mathbf{k}) \\
\mathfrak{\tau}^{21}_{CA\sigma}(\mathbf{k}) & \mathfrak{\tau}^{21}_{CB\sigma}(\mathbf{k}) & \mathfrak{\tau}^{21}_{CC\sigma}(\mathbf{k}) & \mathit{t}^{22}_{CA\sigma}(\mathbf{k}) & \mathit{t}^{22}_{CB\sigma}(\mathbf{k}) & 0
\end{array}
\right)
}
\end{aligned}
\label{hop_mat}
\end{equation} 
}
where the matrix elements, $\tau_{pp'\sigma}^{l\bar{l}}(\mathbf{k})$, result from the HF decomposition of the inter-layer Coulomb repulsion.
\begin{equation}
    \tau_{pp'\sigma}^{l\bar{l}}(\mathbf{k})=-\mathcal{V}_{\perp}P_{p\sigma p'\sigma}^{l\bar{l}}(\mathbf{k}), 
\end{equation}
while $t_{pp'\sigma}^{ll}(\mathbf{k})$ corresponds to the standard momentum space representation of the single particle part of our starting Hamiltonian, corresponding to intra-layer hopping contributions between the $A$, $B$, and $C$ lattice sites of both layers indicated by $l=1,2$ [ cf. Eq. (\ref{ham1})]. 
%{
%\begin{equation}
%\tau_{pp'\sigma}^{ll}(\mathbf{k})=t_{pp'\sigma}^{ll}(\mathbf{k})-%\mathcal{V}_{\perp}P_{lp\sigma\bar{l}p'\sigma}(\mathbf{k}).
%\end{equation}}
All the onsite energies are now contained in the $\mathbf{I}_{l\sigma}$ submatrices of the form
%varepsilon
\renewcommand{\arraystretch}{1}
\begin{equation}
\mathbf{I}_{l\sigma} = 
\left(
\begin{array}{ccc}
\tilde{\varepsilon}_{Al\sigma}-\mu & 0 & 0 \\
0 & \tilde{\varepsilon}_{Bl\sigma}-\mu & 0  \\
0 & 0 & \tilde{\varepsilon}_{Cl\sigma}-\mu
\end{array}
\right),
\end{equation}
 where 
\begin{equation}
%\begin{split}
    \tilde{\varepsilon}_{pl\sigma}=\varepsilon^0_{l}+Un_{pl\bar{\sigma}} + 
\mathcal{V}_{l}\displaystyle\sum_{p'(p'\neq p)}n_{p'l}
%&
+\mathcal{V}_{l\bar{l}}\displaystyle\sum_{p'}n_{p'\bar{l}},
%\end{split}
\end{equation}
with $\mu$ being the chemical potential and $\varepsilon^0_{l=1}=0$, $\varepsilon^0_{l=2}=V+\Delta$. Having the matrix form of our hamiltonian, one can diagonalize it and derive the self-consistent equations for all the mean-fields ($S^z_{lp}=(n_{lp\uparrow}-n_{lp\downarrow})/2$ and $S^{\pm}_{lp}$) as well as the chemical potential in a standard manner. 
Since the system is characterized by significant Coulomb repulsion, the electron-electron correlations may play a role. To take that into account, apart from the HF approximation we additionally apply the Gutzwiller method which is based on the variational wave function of the form
\begin{eqnarray}
|\Psi_{G}\rangle= \prod_{ilp}\mathcal{\hat{P}}_{ilp}|\Psi_{0}\rangle,
\label{jastrow}
\end{eqnarray}
where $|\Psi_0\rangle$ is the non-correlated state of the system and the correlation operator takes the form 
\begin{equation}
    \color{black}\hat{\mathcal{P}}_{ilp}=\mathbf{I}_{ilp}^{\dagger} \mathbf{\Lambda}_{lp} \mathbf{I}_{ilp},
\end{equation}
{where $\mathbf{I}_{ilp}$ is vector composed of states from the local basis,}
\begin{equation}
    \color{black}\mathbf{I}_{ilp}^{\dagger}=\{|\emptyset\rangle_{ilp},|\uparrow\rangle_{ilp},|\downarrow\rangle_{ilp},|\uparrow\downarrow\rangle_{ilp}\},
\end{equation}
{and $\mathbf{I}_{ilp}=(\mathbf{I}_{ilp}^{\dagger})^{\dagger}$, while all the variational parameters are contained in the matrix $\mathbf{\Lambda}_{lp}$ of the form}

\begin{equation}
\color{black}\mathbf{\Lambda}_{lp} = 
\left(
\begin{array}{cccc}
\lambda_{\phi}^{lp}
 & 0 & 0 & 0\\
0 & 
\lambda_{\uparrow\uparrow}^{lp}
 & \lambda_{\uparrow\downarrow}^{lp} & 0  \\
0 & \lambda_{\downarrow\uparrow}^{lp} & \lambda^{lp}_{\downarrow\downarrow} & 0 \\
0 & 0 & 0 & \lambda_{d}^{lp}
\end{array}
\right),
\label{lambda_matrix}
\end{equation}
where we have dropped the $i$ index since we assume that the variational parameters do not depend on the $i$ index but only on the layer index ($l$) and the site index within the supercell ($p$). {This is motivated by the fact that we limit to magnetic or charge modulations with the periodicity corresponding to the chosen supercell}. Additionally, since we do not consider superconductivity here, we have set to zero the variational parameters corresponding to an onsite pairing terms. We consider Hermitian Gutzwiller operators [$(\hat{\mathcal{P}}_{ilp})^{\dagger}=(\hat{\mathcal{P}}_{ilp})$], thus the diagonal terms in Eq. (\ref{lambda_matrix}) are real and $(\lambda_{\downarrow\uparrow})^*=\lambda_{\uparrow\downarrow}$. {The physical meaning of such choice is provided later on.}
{From the technical point of view it is convenient to impose additional constraints to the correlation operator in order to eliminate the trivial local contributions at the inner vertices which arise after the application of Wick's theorem \cite{gutz_con, Gebhard1990,Bunemann_2012}. Namely,}
\begin{eqnarray}
\langle \hat{\mathcal{P}}^{2}_{ilp}\rangle_{0}=1,
\label{cons_1}
\end{eqnarray}
\begin{eqnarray}
\langle{\hat{c}^{\dag}_{i\sigma}\mathcal{\hat{P}}^{2}_{ilp}\hat{c}_{ilp\sigma'}^{}}\rangle_{0}=\langle{\hat{c}^{\dag}_{ilp\sigma}\hat{c}_{ilp\sigma'}^{}}\rangle_{0},
\label{cons_2}
\end{eqnarray}
where $\langle\hat{o}\rangle_0=\langle\Psi_0|\hat{o}|\Psi_0\rangle$.
After taking into account the above constraints, we are left with only one \textit{true} variational parameter per lattice site. In our calculations, we determine $\lambda_d^{lp}$ by minimizing the free energy of the system, and all the remaining $\lambda$'s are determined by using Eqs. (\ref{cons_1}) and (\ref{cons_2}).

In order to derive the expression for the energy expectation value we apply the so-called Gutzwiller approximation which is exact in the limit of infinite dimensions. The resulting formulas are the following
\begin{equation}
\begin{split}
\color{black}\hat{\mathcal{H}'}_{tU} &= \color{black}\sum_{\langle\langle ijpp'\rangle\rangle l\sigma} \tilde{t}^{ll}_{ijpp'\sigma}\;\langle\hat{c}_{i lp \sigma}^{\dagger}\; \hat{c}_{j lp' \sigma}^{}\rangle_0 \\
&\color{black}+\sum_{\langle ijpp'\rangle \sigma} \big(\tilde{t}^{12}_{ijpp'\sigma\bar{\sigma}}\;\langle\hat{c}_{i 1p \sigma}^{\dagger}\; \hat{c}_{j 2p' \bar{\sigma}}^{}\rangle_0+H.c.\big)\\
&\color{black}+(V+\Delta)\sum_{ip}\langle\hat{n}_{ip\;l=2}\rangle_0,
\end{split}
\label{ham1_again}
\end{equation}
\begin{eqnarray}
\langle\mathcal{\hat{H}}_U\rangle_G  = U\sum_{ilp}(\lambda^{lp}_{d})^{2}\langle \hat{n}_{ilp\uparrow} \hat{n}_{ilp\downarrow} \rangle_{0},
\label{renorm_ham_1}
\end{eqnarray}
where $\langle \hat{o} \rangle_G$=$\langle\Psi_G| \hat{o} |\Psi_G\rangle/\langle\Psi_G|\Psi_G\rangle$ and
\begin{equation}
\begin{split}
\widetilde{t}^{ll'}_{ijpp'\sigma\sigma'} &= {t}_{ijpp'\sigma\sigma'}^{ll'}{q}_{ilp\sigma\sigma}^{*}{q}_{jl'p'\sigma'\sigma'}^{}\\
&+{t}_{ijpp'\bar{\sigma}\bar{\sigma'}}^{ll'}{q}_{ilp\bar{\sigma}\bar{\sigma'}}^{*}{q}_{jl'p'\bar{\sigma}\bar{\sigma'}}^{}.
\label{renorm_t}
\end{split}
\end{equation}
In the above Equation the diagonal and off-diagonal renormalization hopping factors have the form,
\begin{equation}
\begin{split}
{q}_{lp\sigma \sigma}^{*}&= \lambda^{lp}_{\sigma\sigma}\lambda^{lp}_{\phi}
-n_{lp\bar{\sigma}}\big( \lambda^{lp}_{\sigma\sigma}\lambda^{lp}_{\phi}-\lambda^{lp}_{d}\lambda^{lp}_{\bar{\sigma}\bar{\sigma}}\big)\\
&+n_{lp\bar{\sigma}\sigma}\big(\lambda^{lp}_{d}({\lambda^{lp}_{\sigma \bar{\sigma}})}^{*}+({\lambda^{lp}_{\sigma \bar{\sigma}})}^{*}\lambda^{lp}_{\phi} \big),
\end{split}
\label{q_renormalization_1}
\end{equation}

\begin{equation}
\begin{split}
{q}_{lp\bar{\sigma}\sigma }^{*}&=\lambda^{lp}_{\sigma \bar{\sigma}}\lambda^{lp}_{\phi}-n_{lp\bar{\sigma}}\big(\lambda^{lp}_{\sigma \bar{\sigma}}\lambda^{lp}_{\phi}+\lambda^{lp}_{d}\lambda^{lp}_{\sigma \bar{\sigma}}\big)\\
&-n_{lp\bar{\sigma}\sigma}\big(\lambda^{lp}_{d}\lambda^{lp}_{\sigma\sigma}-\lambda^{lp}_{\bar{\sigma}\bar{\sigma}}\lambda^{lp}_{\phi}\big),
\end{split}
\label{q_renormalization_2}
\end{equation}
where $\bar{\sigma}=\uparrow$ ($\bar{\sigma}=\downarrow$) when $\sigma=\downarrow$ ($\sigma=\uparrow$). For the calculations carried out with the use of the Gutzwiller approximation, we restrict ourselves to the situation with onsite interactions only, therefore, we do not provide the expectation value of the $\hat{\mathcal{H}}_{\mathcal{V}}$ part of the Hamiltonian in the $|\Psi_G\rangle$ state. 

As one can see from Eqs. (\ref{ham1_again}) and (\ref{renorm_ham_1}), the expectation value of the original Hamiltonian in the correlated state can be expressed in terms of the expectation values in the non-correlated state. Therefore, one can construct an effective Hamiltonian $\hat{\mathcal{H}'}_{tU}$ such that $\langle\hat{\mathcal{H}'}_{tU}\rangle_0=\langle\mathcal{H}_{tU}\rangle_G$ within the Gutzwiller approximation. {The explicit form of the effective Hamiltonian is provided below for the sake of completeness}
\begin{equation}
    \begin{split}
\color{black}\hat{\mathcal{H}'}_{tU} &= \color{black}\sum_{\langle\langle ijpp'\rangle\rangle l\sigma} \tilde{t}^{ll}_{ijpp'\sigma}\;\hat{c}_{i lp \sigma}^{\dagger}\; \hat{c}_{j lp' \sigma}^{} \\
&\color{black}+\sum_{\langle ijpp'\rangle \sigma} \big(\tilde{t}^{12}_{ijpp'\sigma\bar{\sigma}}\;\hat{c}_{i 1p \sigma}^{\dagger}\; \hat{c}_{j 2p' \bar{\sigma}}^{}+H.c.\big)\\
&\color{black}+(V+\Delta)\sum_{ip}\hat{n}_{ip\;l=2}\\
&\color{black}+U\sum_{ilp}(\lambda^{lp}_{d})^{2} \hat{n}_{ilp\uparrow} \hat{n}_{ilp\downarrow}.
\end{split}
\label{eq:hamiltonian_effective}
\end{equation}

{The fact that $\hat{\mathcal{H}'}_{tU}$ is hermitian is a consequence of the choice of hermitian Gutzwiller operator as mentioned earlier.} In such an approach, Hamiltonian $\hat{\mathcal{H}'}_{tU}$ can be transformed to the momentum space and treated within the mean-field method with the correlation effects taken into account via the renormalization factors given by Eqs. (\ref{q_renormalization_1}), (\ref{q_renormalization_2}), and $\lambda^{lp}_d$. {However, the procedure of solving the self-consistent equations has to be coupled with the minimization of the energy of the system over the variational parameters. From the technical point of view this can be done by supplementing the set of self consistent equations with the following} 
\begin{equation}
    \color{black}\frac{\partial \langle \hat{\mathcal{H}}\rangle_G}{\partial\lambda_{d}^{lp}}=0,
\end{equation}
{where in our case the analytical derivative is replaced by a numerical one.}
It should be noted that here we are mainly focused on the situation with one hole per moir\'{e} lattice site, when the lower band is completely filled and the upper one is half-filled. In general it is expected that the renormalization is significant close to the half-filled situation. While moving away from half-filling the renormalization parameters are becoming close to one. Therefore, for the sake of simplicity, we only take into account the renormalization in the MoTe$_2$ band.

The restricting conditions Eq. (\ref{cons_1}) and (\ref{cons_2}) are used to generate the necessary equations, which are solved numerically using a hybrid subroutine from the MINPACK library \cite{minpack}, where \(\lambda_d\) is varied self-consistently in our calculations. It uses finite difference approximation of the Jacobian to estimate the Jacobian matrix when solving systems of nonlinear equations.

%%%%%%%%%%%%%%%%%%%%%%%%%%%%%%%%%%%%%%%%%%%%%%%%%%%%%%%%%%%%%%%%%%%%%%%%%%%%%%%%%%%%%

%\section{Results}
\section{\label{sec:level3}Results}

Within this Section we show the results of our calculations divided to two subsection. In subsection A we focus on the analysis of the topological and magnetic features of the system limiting to the onsite Coulomb repulsion only, whereas, the influence of the intersite repulsion as well the analysis of the charge ordered states is deferred to subsection B. For the sake of clarity while analyzing our results we replace the $l=1$ and $l=2$ indices which correspond to MoTe$_2$ and WSe$_2$ layers with $M$ and $W$ subscripts, respectively. 

\subsection{QAHI state formation}
We first apply the Hartree-Fock approximation in order to analyze the formation of a non-trivial topological state within the Hubbard model of $MoTe_2/WSe_2$ heterobilayer for one hole per moir\'{e} unit cell. At this filling, the QAHI behavior has been reported experimentally indicating the appearance of conductive edge states and an insulating bulk of the sample \cite{mw2}. In our analysis, we use the electron language, for which $n_{tot}=3$ is equivalent to one hole per moir\'{e} unit cell and corresponds to completely filled lower band and half-filled upper band (cf. Fig. \ref{two_bands}). For the sake of simplicity, we initially focus on the on-site Coulomb repulsion, taking $\mathcal{V}_{\perp}=\mathcal{V}_{\parallel}=0$, and introduce the following relation $U_M=U_W\equiv U$. Following the recent experimental report \cite{Zhao2023} we aim at reconstructing the band structure corresponding to a charge transfer insulator induced by substantial Hubbard $U$. Hence, we set a relatively large value of $U/|t_1|=21$, which is comparable to the one taken in Ref. \onlinecite{Xi_2024}. 

\begin{figure}
\includegraphics[width=0.95\linewidth, height=8.5cm]{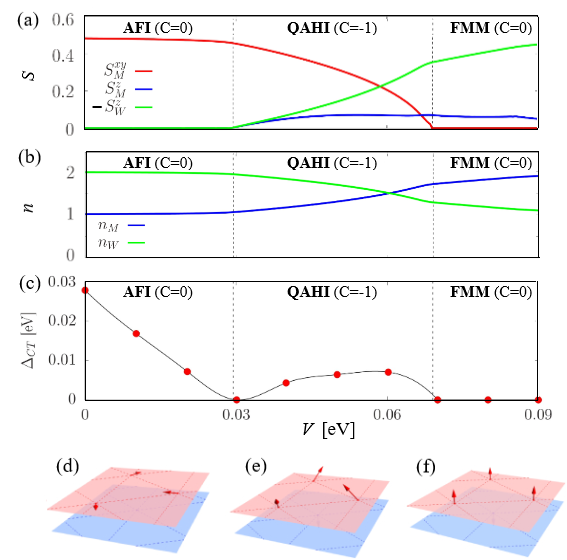}
\caption{(a) The in-plane magnetization corresponding to 120$^\circ$ AF ordering at the MoTe$_2$ layer ($S^{xy}_M$), together with the out-of-plane magnetizations in both layers ($S^z_M$ and $-S^z_W$), all as functions of the displacement field. The corresponding in-plane magnetization at the WSe$_2$ layer is zero in the whole considered range of $V$. The sequence of states that appear with increasing $V$ is: antiferromagnetic charge transfer insulator (AFI); quantum anomalous Hall insulator (QAHI) with non-zero Chern number $C=-1$; ferrimagnetic metal (FMM). 
(b) The number of electrons per moir\'{e} lattice site in both layers as a function of displacement field. (c) The band gap between the WSe$_2$ band and the upper MoTe$_2$ subband. In (d), (e), and (f) we show the graphical representation of the magnetic ordering in the AFI, QAHI, and FMM phases. The blue and red surfaces correspond to the WSe$_2$ and MoTe$_2$ layers. The results have been obtained for $U/|t_1|=21$ and $n_{tot}=3$, which is equivalent to one hole per moir\'{e} unit cell.}
\label{mean_n}
\end{figure}

The calculated evolution of the magnetically ordered states with increasing displacement field is shown in Fig. \ref{mean_n}(a). Initially, for small values of $V$ an in-plane 120$^\circ$ antiferromagnetic ordering is realized at the MoTe$_2$ layer with the in-plane magnetization $S^{xy}_M$ playing the role of the order parameter. No magnetic ordering appears at the WSe$_2$ layer since its band is fully filled. {At some critical value of the displacement field a second order phase transition takes place and the out-of-plane magnetization in both layers become non-zero ($S_M^z> 0$ and $S_W^z<0$), leading to a canted AF order at the MoTe$_2$ layer}. According to our analysis, this region corresponds to a Chern number $|C|=1$ indicating a quantum anomalous Hall insulating (QAHI) state. The details of the Chern number calculations, together with the Berry curvature maps, are provided in Appendix A.
\begin{figure}[t]
\centering
\includegraphics[width=0.92\linewidth, height=11cm]{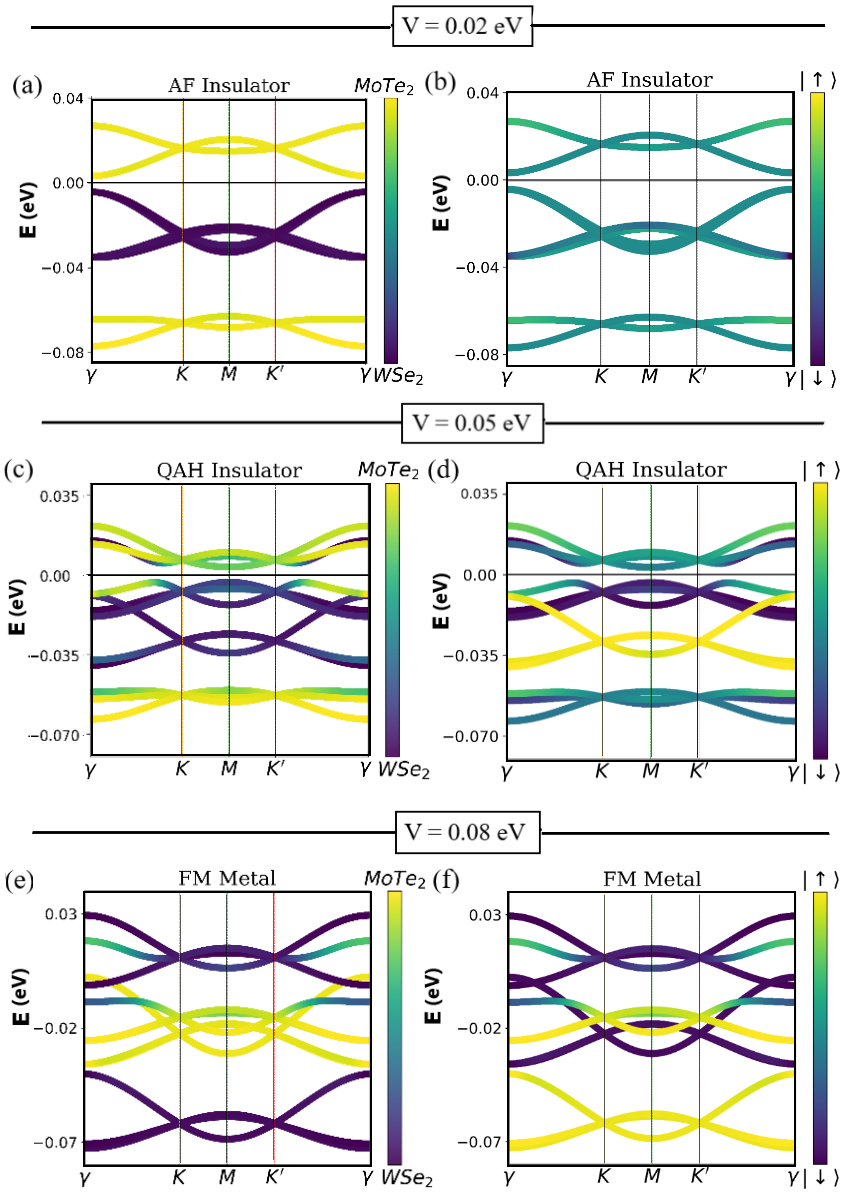}
\caption{The band structure for the three representative values of the displacement field which correspond to the AF insulator (a,b), quantum anomalous Hall insulator (c,d), and ferrimagnetic metal (e,f). The colored scale in the first (second) column indicates the layer (spin) contribution to the resulting state. The results have been obtained for the same model parameters as in Fig.~\ref{mean_n}.}
\label{six_bands}
\end{figure}
Finally, for high enough values of $V$ a second transition appears where the in-plane AF ordering is destroyed and only out-of-plane magnetization survives, leading to a ferrimagnetic metallic state (FMM), in which the magnetic moments of the two layers are oriented antiparallel but the one corresponding to the WSe$_2$ layer dominates. Additionally, in Fig. \ref{mean_n}(b) we show the calculated number of electrons per MoTe$_2$ and WSe$_2$ moir\'{e} orbitals ($n_M$, $n_W$). As one can see in the Figure, in the AFI state we start from the half-filled MoTe$_2$ band ($n_M\approx 1$) and completely filled WSe$_2$ band ($n_W\approx 2$). After entering the QAHI region, due to band mixing, the electrons are starting to be transferred from the WSe$_2$ layer to the MoTe$_2$ layer leading to a more homogeneous distribution of charge between the layers. However, close to the transition to the FMM state the situation becomes inverted. It should be noted that the appearance of magnetic ordering with non-trivial topological features characterized by $|C|=1$, has been reported quite recently and discussed in the context of MoTe$_2$/WSe$_2$ in Ref. \onlinecite{mw2}. {However, in that analysis, a model different from ours has been considered with purely real and spin-conserving interlayer hoppings as well as different phase factors in the intra-layer hoppings.}
\begin{figure}[t]
\centering
\includegraphics[width=0.9\linewidth, height=6cm]{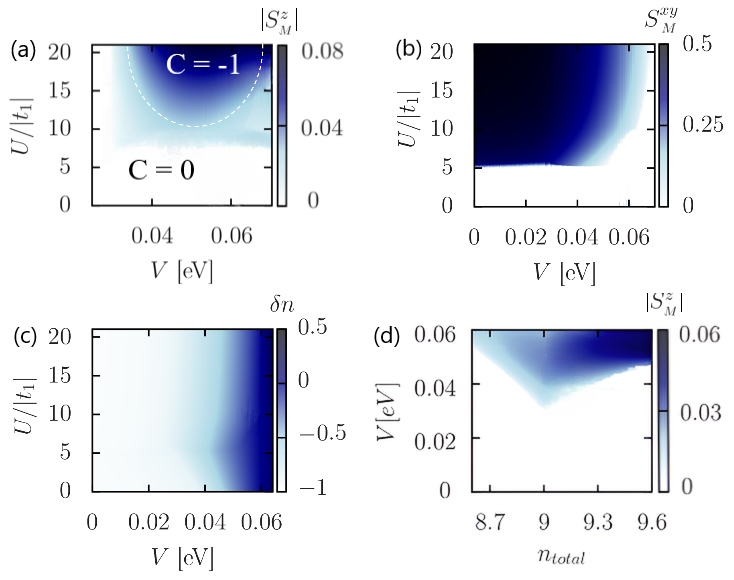}
\caption{The out-of-plane magnetization (a) and in-plane magnetization (b) in the MoTe$_2$ layer as well as $\delta n=n_M-n_{WS}$ (c), all as functions of the displacement field ($V$) and onsite Coulomb repulsion ($U$) for half-filling of the upper band ($n_\text{tot}=3$). In (d) we show the out-of-plane magnetization for selected values of $U/|t_1|=10.7$ as a function of displacement field and total number of electrons.}
\label{4_images}
\end{figure}

In order to gain some more insight in the origin of the obtained sequence of phases, in Fig. \ref{six_bands}, we show the band structure for the three representative values of the displacement field corresponding to the three regimes seen in Fig. \ref{mean_n}. The colored scale indicates the layer and spin state contribution in a given band. It should be noted that we plot the bands along the high symmetry points of the reduced mini Brillouin zone. The reduction is due to the introduction of the supercell which allows for the AF alignment in our calculations (cf. Sec. II). As one can see in the AFI  state the upper MoTe$_2$ band seen in Fig. \ref{two_bands} has spit into two subbands with the bottom one being located below the WSe$_2$ band. For $n_{tot}=3$ the upper MoTe$_2$ subband is completely empty and separated by an energy gap from the fully occupied WSe$_2$ band and the lower MoTe$_2$ subband. Such band structure is consistent with the charge transfer insulating scenario. As the displacement field is increased the WSe$_2$ band moves up, which from some point induces band mixing and emergence of the spin splitting due to the out-of-plane magnetization discussed earlier. In this state the gap is still open, but now the calculated Chern number of the upper MoTe$_2$ subband is $|C|=1$, which induces the QAHI behavior.
{
While the Chern number is typically calculated for an individual band, the presence of degeneracies at band crossings in our case means that at each wavevector, we obtain a three-dimensional subspace of states. Consequently, the Berry connection becomes a non-Abelian \(3\times3\) matrix, often termed as the Berry rotation matrix that governs state mixing. This induces non-commuting connection elements and extends topological classification from the first Chern number to the non-Abelian second Chern number associated with the matrix-valued Berry gauge field in momentum space. For more details about the case of  discretization in the Brillouin zone, the Appendix A is suggested for a quick revision.}

It should be noted that due to the AF alignment in the QAHI state the spin lowering ($S^-$) and raising ($S^+$) expectation values are non-zero which together with the band mixing term leads to a situation in which a hybridization of states from different layers and different valleys is allowed near the Fermi level. For relatively large values of the displacement field the AF ordering is destroyed and the band splitting of the MoTe$_2$ band disappears which results in gap closing and one is left with an out-of-plane ferrimagnetic metallic state.

The results shown in Figs. \ref{mean_n} and \ref{six_bands} agree qualitatively with the available experimental data in the following aspects. First, with increasing displacement field the system undergoes, a transition from the topologically trivial insulting to the QAH insulating state and then, a transition from the QAH to the metallic state appears\cite{Tao2024, mw_exp_1}. Secondly, the QAH state possesses a spontaneous spin-polarization and the holes are distributed in both layers\cite{Tao2024}. Moreover, in the QAH state, hybridization of states from different layers and different valleys is allowed near the Fermi level. One significant difference between our calculations and the experiment is the fact that we see a continuous gap closing while approaching the QAHI state, whereas in the experiments the gap seems to remain finite at the transition\cite{mw_exp_1}. {With respect to that, recently it has been theoretically proposed that due to Coulomb interactions a Nematic Excitonic Insulating state may appear, which breaks the $C_{3}$ symmetry and preempts the phase transition to the Chern insulator\cite{r4}. As shown, such a precursor state might keep the gap open before entering the QAH state in agreement with the experimental report mentioned.}

\begin{figure}[t]
\includegraphics[width=0.95\linewidth, height=3.5 cm]{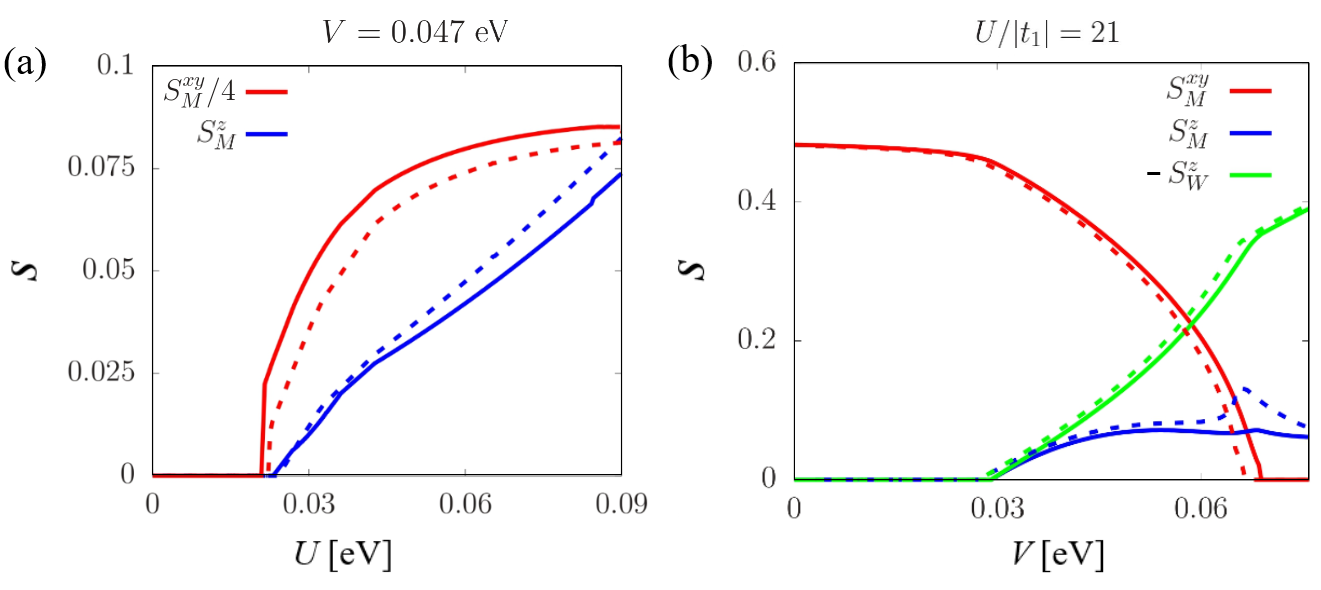}
\caption{(a) The out-of-plane and in-plane magnetization for the MoTe$_2$ layer as a function of $U$ for $V=0.047$ eV. (b) The out-of-plane and in-plane magnetizations as a function of $V$ for $U/|t_1|=21$. The solid lines correspond to the Hartree-Fock calculations and the dashed lines represents the results obtained with the use of the Gutzwiller approximation.}
\label{gutz1}
\end{figure}

For the sake of completeness in Fig. \ref{4_images} we show the in-plane and out-of-plane magnetization in the MoTe$_2$ layer as functions of onsite Coulomb repulsion and displacement field. Since the magnetically ordered states are spontaneously induced by electron interactions, some minimal value of $U$ is needed to stabilize them. Note that the non-trivial topological state on the $(U,V)$-plane is obtained for the area in which both $S^z_M\neq 0$ and $S^{xy}_M\neq 0$ [marked by the dashed white line in (a)]. As one can see, the range of displacement fields in which the QAHI state is stable, is getting narrower with decreasing $U$. The minimal value for the appearance of QAHI is $U\sim W$, where $W$ is the bare bandwidth of the upper band, which is $W\approx 9|t_1|$ in our case. Additionally, in (d) we show how the out-of-plane magnetization on the MoTe$_2$ layer changes as one moves away from the half-filled situation. The results shown in (d) have been obtained for $U/|t_1|=10.7$, which still leads to the appearance of the QAH but due to the relatively smaller value of $U$ allows for better numerical convergence. One should note, that in this case only $n_{tot}=3$ corresponds to an insulating state, since only then the fully occupied subbands/bands are separated by an energy gap from the fully empty subband (cf. Fig. \ref{six_bands}).

%%%%%%%%%%%%%%%%%%%%%%%%%%%%%%%%%%%%%%%%%%%%%%%%%%%%%%%%%%%%%%%%%%%%%%%%%%%%%%%%%%%%
The comparison between the GA and HF methods is provided in Fig. \ref{gutz1}. As one can see in the considered case, both methods give similar results. It should be noted that within the Gutzwiller approximation the wave function is renormalized the most for the case of one electron per lattice site since in such case electron hopping leads to creation of double occupancy which in turn corresponds to significant increase of system energy due to strong onsite Coulomb repulsion. For a single-band model half-filling always corresponds to such a scenario. However, in a two-band model analyzed here, where band mixing appears, even for half-filling of the upper band ($n_{tot}=3$), the electrons can be distributed in the layers in such a manner that $n_W\neq 1$ and $n_M\neq 1$ (cf. Fig. \ref{mean_n}). That diminishes the effect of the correlation operator, most probably leading to similarities between the HF and GA calculations. Also, for $V=0$ when the band mixing is absent, we are dealing with nearly saturated 120$^{\circ}$ AF state and again both in GA and HF are close.

%%%%%%%%%%%%%%%%%%%%%%%%%%%%%%%%%%%%%%%%%%%%%%%%%%%%%%%%%%%%%%%%%
\subsection{Influence of the intersite Coulomb repulsion}
\begin{figure}[t]
\includegraphics[width=0.99\linewidth, height=7.5cm]{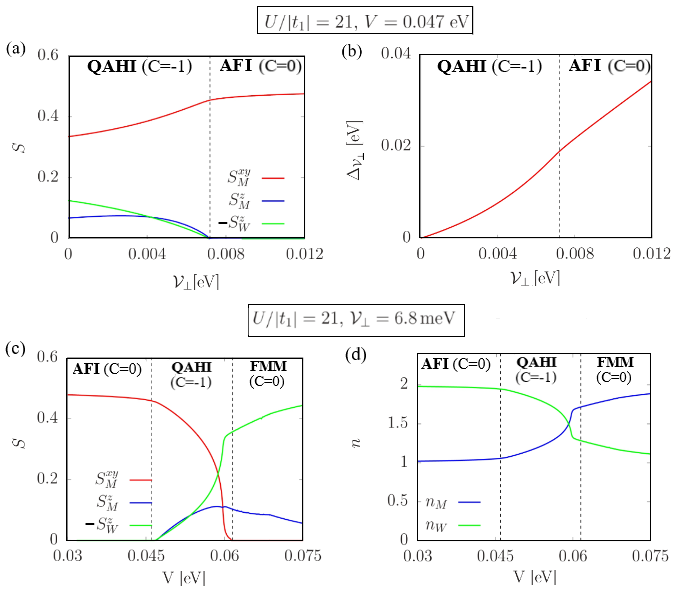}
\caption{(a) The in-plane magnetization corresponding to 120$^{\circ}$ AF ordering at the MoTe$_2$ layer ($S^{xy}_{M}$), along with the out-of-plane magnetizations in both layers ($S_M^z$ and $S_W^z$), all as functions of the interlayer intersite Coulomb repulsion constant, $\mathcal{V}_{\perp}$.
(b) The difference between the MoTe$_2$ and WSe$_2$ onsite energies resulting from the Hartree-Fock decomposition of the $\mathcal{V}_{\perp}$-term [cf. Eq. (\ref{delta_onsite_perp})]. Note that by increasing $\mathcal{V}_{\perp}$, one obtains a similar effect to that resulting from decreasing the displacement field (cf. Fig. \ref{mean_n}).
{(c) In-plane magnetization at the MoTe$_2$ layer ($S^{xy}_M$) and out-of-plane magnetizations in both layers ($S^z_M$, $S^z_W$) as functions of displacement field across the three phases. (d) Variation of the average particle number per lattice site in the MoT$e_2$ and WS$e_2$ layers as a function of the displacement field.}
}
\label{inter}
\end{figure}

Here we analyze the influence of the intersite Coulomb repulsion terms. We first set the displacement field to the value of $V=0.047$ eV, which corresponds to the stability of the QAHI, and analyze the evolution of this state with increasing interlayer Coulomb term $\sim\mathcal{V}_{\perp}$. In Fig. \ref{inter} we show the $\mathcal{V}_{\perp}$-dependence of both in-plane and out-of-plane magnetizations together with the value of the Chern number corresponding to particular considered regimes. It should be noted that within the HF approach, the $\mathcal{V}_{\perp}$-term leads to an effective onsite energies at the two layers [cf. Eq. (\ref{ham_V_HF})]. The difference between the MoTe$_2$ and WSe$_2$ onsite energies resulting from the considered effect is shown in Fig. \ref{inter}(b) and can be expressed in the following manner
\begin{equation}
    \Delta_{\mathcal{V}_{\perp}}=3\mathcal{V}_{\perp}(n_W-n_M).
    \label{delta_onsite_perp}
\end{equation}
With increasing $\Delta_{\mathcal{V}_{\perp}}$ one actually increases the charge transfer gap between the upper MoTe$_2$ subband and the WSe$_2$ band (cf. Fig. \ref{six_bands}). Such effect is equivalent to decreasing the displacement field shown in the previous part of our analysis (cf. Fig. \ref{mean_n}). Consequently, by enhancing $\mathcal{V}_{\perp}$ one eventually can destroy the QAHI state by suppressing the band mixing. Additionally, in \ref{inter} (c) and (d) we show the evolution of the system with changing displacement field similarly as in Fig. \ref{mean_n} but now for nonzero $\mathcal{V}_{\perp}$. As one can see after the inclusion of the $\mathcal{V}_{\perp}$-term, the critical values of the displacement field, at which transitions between different phases appear, have changed. In particular the QAHI stability regime has narrowed down. Nevertheless, from the qualitative point of view the behavior is similar as before.

%%%%%%%%%%%%%%%%%%%%%%%%%%%%%%%%%%%%%%%%%%%%%%%%%%%%%%%%%%
\begin{figure*}[t]
\centering
\includegraphics[width=1\linewidth, height=4cm]{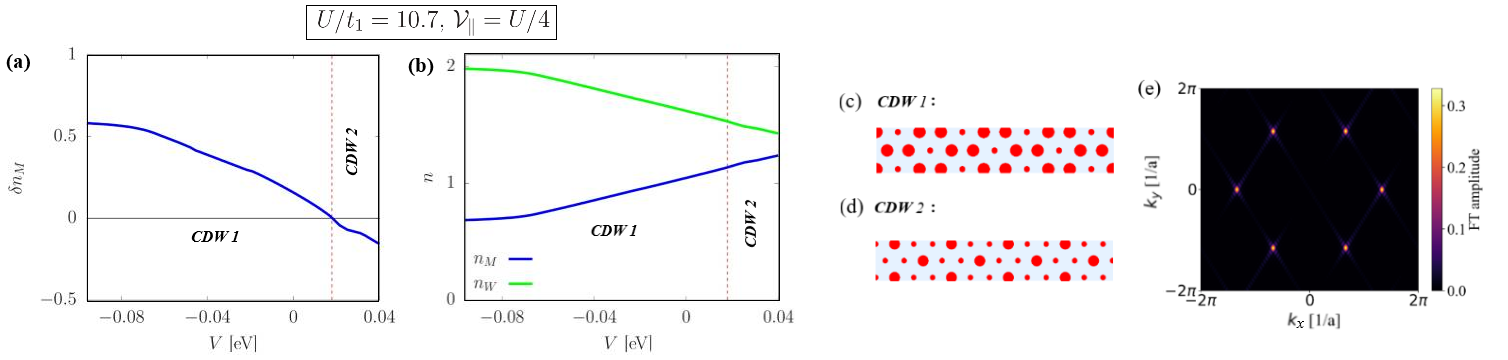}
\caption{(a) {Charge modulation amplitude at the MoTe$_2$ layer [cf. Eq. (\ref{CDW_amplitude})] as a function of displacement field \(V\).} (b) Average number of particles per lattice site at the MoTe$_2$ layer and WSe$_2$ layer as a function of displacement field. (c) and (d) show a visual representation of the charge ordered states where red dots with larger radius represent larger electron concentration occupying given lattice site. (c) correspond to the lower $V$ values where a honeycomb lattice composed of the increased concentration lattice sites is present while (d) corresponds to higher $V$ values where an inverse situation is realized. In (c) and (d) we show only the lattice sites of the MoTe$_2$ layer since only there the charge ordering takes place. {(e) The Fourier transform of the charge pattern which corresponds both to the situation represented by (c) and (d). Here we have subtracted the peaks resulting from the presence of the crystal lattice itself, so that the resulting amplitude corresponds only to the CDW formation.}}
\label{intra}
\end{figure*}

Next we include the intralayer intersite Coulomb repulsion and focus on the scenario of $n_{tot}=2\frac{2}{3}$ in order to study the formation of the charge ordered phase. In this part of the analysis, we have decreased the value of onsite Coulomb repulsion to $U\approx W$ which leads to relatively better convergence of the numerical procedure. Following Ref. \onlinecite{Xi_2024} we set $\mathcal{V}_{\parallel}=U/4$ and initially consider a relatively large negative bias voltage to separate significantly the MoTe$_2$ and WSe$_2$ bands leading to a fractional filling of $n_M=2/3$ and $n_W=2$, as shown in Fig. \ref {intra} (b). 
We find a spontaneous formation of a charge modulation resulting from carrier transfer from the $C$ lattice site to the $A$ and $B$ lattice sites within the supercell [cf. Fig. \ref{af_lattice}]. In such case the carrier concentration in the three lattice sites of the MoTe$_2$ layer can be expressed in the following manner
\begin{equation}
    \color{black}n_A=n_B=\bar{n}+\frac{\delta n_{M}}{2},\quad n_C=\bar{n}-\delta n_M,
    \label{CDW_amplitude}
\end{equation}
{where $\delta n_M$ and $\bar{n}$ are the amplitude of the modulations and the base value, respectively.} In Fig. \ref{intra}(a) we show the $\delta n_M$ parameter as functions of the displacement field. As one can see for large negative $V$, we have $\delta n_M\approx 2/3$ meaning that the $C$ lattice site of the MoTe$_2$ layer becomes nearly vacant, with two remaining sites ($A$ and $B$) being singly occupied. Such charge distribution minimizes the interaction energy resulting from the $\mathcal{V}_{\parallel}$-term. Similarly as before, by increasing the displacement field we change the distribution of carriers among the two layers [cf. Fig. \ref{intra}(b)]. This weakens the amplitude of the charge modulation since with increasing $V$ we are moving away from the fractional filling of $n_M=2/3$. Consequently, at some point, $\delta n_M$ changes sign and the situation becomes inverted, meaning that a transition appears from a honeycomb structure of increased carrier concentration, shown in Fig. \ref{intra}(c), to a honeycomb structure of decreased carrier concentration shown as in Fig. \ref{intra}(d). {Additionally, in Fig. \ref{intra} (e) we show the Fourier transform of the charge modulation where, for the sake of clarity, we have subtracted the peaks resulting from the presence of the triangular crystal lattice itself. In such a manner, only the contribution resulting from the obtained charge modulation is visible. The resulting six peaks are located at the corners of a triangular lattice Brillouin zone and correspond to the charge density wave modulation vectors realized in both $\delta n_M>0$ and $\delta n_M<0$ case.}
Note that different forms of charge ordered states have been experimentally evidenced in TMD based moir\'{e} systems at fractional fillings\cite{Regan2020,Li2021,Mak2021CDW,Feng2024}. In particular in WSe$_2$/WS$_2$, a honeycomb charge pattern, similar to the one analyzed here, has been imaged by using a non-invasive STM spectroscopy technique\cite{Li2021}. The WSe$_2$/WS$_2$ system should correspond a very similar single particle model as the one considered here for the case of the MoTe$_2$/WSe$_2$ as shown in Ref. \onlinecite{louk}.  
{Finally, it should be noted that charge ordered states can also show different forms of magnetic ordering. Such effect has been recently analyzed theoretically for the WSe$_2$/WS$_2$ heterobilayer\cite{Rademaker_Kagome,Biborski_2025}. For the sake of completeness, we have analyzed the magnetic ordering of the obtained charge pattern for the case of relatively strong negative bias voltage for which the CDW state is equivalent to a honeycomb structure of singly occupied sites (cf. Fig. \ref{intra}). According to our study the obtained state shows a canted ferromagnetic ordering in the $x-y$ plane which is consistent with the result obtained by us recently in Ref. \cite{Biborski_2025} for WSe$_2$/WS$_2$ heterobilayer. Note that by carrying out a gauge transformation on the obtained magnetic state one obtains a standard staggered AF ordering, as expected, for a honeycomb structure of singly occupied sites.}

\section{Conclusions}
We applied the effective two-band moir\'{e} Hubbard model to analyze the interplay between electron-electron interactions and the topological features of MoTe$_2$/WSe$_2$ heterobilayer. According to our calculations, at half filling ($n_{tot}=3$) and for large enough values of the onsite Coulomb repulsion the upper MoTe$_2$ band splits into two subbands creating an in-plane 120$^{\circ}$ AF charge transfer insulator. However, by increasing the displacement field one decreases the charge transfer gap leading to band inversion as well as the appearance of an out-of-plane magnetization. These two effects result in a formation of charge transfer QAH insulating state. By a further increase of the displacement field, one suppresses the in-plane AF insulating state leading to a out-of-plane ferrimagnetic metallic state. Such sequence of phases together with the fact that holes are distributed in both layers, agrees qualitatively with the available experimental data \cite{mw_exp_1,Tao2024}. Another aspect which is seen both here and in the experiments is that the QAHI state possesses a spontaneous spin-polarization in both layers and the holes are distributed in both layers\cite{Tao2024}.

Additionally, we have analyzed the influence of the inter-site Coulomb repulsion terms. In particular, as we show the intersite interlayer Coulomb term ($\sim \mathcal{V}_{\perp}$) leads to a similar effect as the one induced by decreasing the displacement field. As such it might have a destructive influence on the QAHI state. Such effect can be understood in terms of effective onsite energies tuned by the $\mathcal{V}_{\perp}$ after the Hartree-Fock decomposition. Namely by increasing $\mathcal{V}_{\perp}$ one increases the charge transfer gap removing the band inversion which is crucial for the QAHI state formation. 

After the inclusion of the intersite intralayer Coulomb repulsion term ($\sim \mathcal{V}_{\parallel}$), we have considered the fractional filling of the upper band, in order to analyze the possibility of charge ordering. According to our calculations the charge density wave state realizing a pattern of honeycomb lattice sites of increased electron concentration at the MoTe$_2$ layer, is realized in the regime of low displacement fields. However, since by displacement field one changes the distribution of charge between the layers the CDW amplitude is decreased with increasing $V$. At some point a inverse situation appears of a pattern of honeycomb lattice sites with decreased electron concentration at the MoTe$_2$ layer.

For the sake of completeness we have also supplemented our Hartree-Fock result with those stemming from the Gutzwiller approximation method. According to this part of the analysis the changes seems to be quantitative with stronger tendency towards magnetic ordering appearing for the GA solution. However, from the qualitative point of view both methods lead to similar physical picture of the considered model.

The code that was written to perform the numerical calculations and the data behind the figures are available in the open repository\cite{zegrodnik_michal_2024_zenodo}.

\begin{acknowledgments}
 This  research was supported by National Science Centre, Poland (NCN) according to Decision No. 2021/42/E/ST3/00128. We gratefully acknowledge Polish high-performance computing infrastructure PLGrid (HPC Center: ACK Cyfronet AGH) for providing computer facilities and support within computational grant no. PLG/2024/017372. For the purpose of Open Access, the author has applied a CC-BY public copyright licence to any Author Accepted Manuscript (AAM) version arising from this submission.

\end{acknowledgments}

\appendix

\section{Second Chern Number}
Here, we present a comprehensive numerical methodology which has been applied to accurately compute the second Chern number characterizing the considered system.

{Chern number is typically calculated for an individual well separated band. Here, due to the introduction of supercell consisting of six moir\'{e} orbitals (cf. Fig. \ref{af_lattice}), the resulting band structure consists of nine dispersion relation branches in the folded mini Brillouin zone (cf. Fig. \ref{six_bands}). For $n_{tot}=3$ the upper three branches correspond to empty states, which in the QAHI state are separated by a band gap from the remaining six occupied bands. In general, while calculating the Chern number one takes into account the occupied bands. However, equivalently a transition to the hole language can be carried out and than we only need to consider the three upper bands. Nevertheless, there still are crossings between the three bands; therefore, while calculating the Chern number we have to consider a three-dimensional subspace formed by the hole occupied states which, in general, can be degenerate. The following approach leverages some basics of numerical approximations in lattice gauge theory. Instead of a single state, we will consider a subspace spanned by a set of three degenerate states (\(\alpha,\beta \in [1,3]\)) with the same quantum number \(n\).}

{A three dimensional subspace
\begin{equation}
H_{n} = \left\{ \sum_{\beta=1}^{3} c_\beta \ket{\psi_{n_\beta \textbf{k}}} ,\quad c_\beta \in \mathbb{C}\right\},
\end{equation}
consists of the orthonormal set of eigenvectors \(\ket{\psi_{n_\beta \textbf{k}}}\) which locally spans \( H_{n} \), satisfies
\begin{equation}
\braket{\psi_{n_\alpha \textbf{k}} | \psi_{n_\beta \textbf{k}}} = \delta_{\alpha \beta}.
\end{equation}}
For the adiabatic evolution of a quantum state, the superimposed state is obtained from a unitary transformation of the original states,

\begin{equation}
|\Psi_{n_{\alpha}\textbf{k}}\rangle= \sum_{\beta=1}^{3}{U}^{n}_{\alpha\beta \textbf{k}}|\psi_{n_{\beta}\mathbf{k}}\rangle,
\end{equation}
where, the unitary matrix \cite{wzee} contains all the connection terms along a closed contour \(\mathcal{C}\) given by, 
\begin{equation}
{U}^{n}_{\mathbf{k}}(t)= \mathcal{P}e^{i \oint_{\mathcal{C}}{{d\mathbf{k}}.\mathbf{\mathcal{A}}^{n}(\mathbf{k})}}.
\end{equation}
This is the \(U(3)\) gauge transformation applied to describe transformation properties of gauge potential and quantum state during it's evolution.

The path ordering is crucial so that the non-Abelian gauge potential can remain non-commutative, thus following the properties of a matrix with each element given by, 
\begin{equation}
\mathcal{A}^{n}_{ij}(\mathbf{k})= \langle{\psi_{n_{i}\mathbf{k}}|\bm{\Delta_{\mathbf{k}}}\psi_{n_{j}\mathbf{k}}}\rangle.
\end{equation}
{
In the continuum \(\textbf{k}\)-space, the Berry curvature takes the form  
\begin{equation}
\Omega_{\mu \nu} = \partial_{\mu} \mathcal{A}_{\nu} - \partial_{\nu} \mathcal{A}_{\mu} + [\mathcal{A}_{\mu}, \mathcal{A}_{\nu}],
\label{curv_1}
\end{equation}  
where the gauge field \( \mathcal{A}_{\mu} \) acts as a scalar potential when the components commute, \( [\mathcal{A}_{\mu}, \mathcal{A}_{\nu}] = 0 \), but takes on a matrix structure when they do not, leading to \( [\mathcal{A}_{\mu}, \mathcal{A}_{\nu}] \neq 0 \).
}

To recover the discrete formula, we use the method used by Fukui-Hatsugai-Suzuki \cite{fukui}. We need to define a \(U(1)\) link variable for parallel transport between any two neighboring points \(\mathbf{k}\) and \(\mathbf{k}+\mathbf{k}_{\Gamma}\) in the base manifold, 
\begin{equation}
{\mathcal{U}^{n}_{\Gamma}}_{|\alpha\beta}(\mathbf{k})= \frac{1}{\mathcal{N}}\langle{\psi_{n_{\alpha}\mathbf{k}}|\psi_{n_{\beta}\mathbf{k}+\mathbf{k}_{\Gamma}}}\rangle ,
\end{equation}
where, \(\Gamma=\{x,y\}\) and \(\mathcal{N}\) is the normalization constant. 
{
In the Abelian case, the phase factors \( \mathcal{U}_{\mu}(\textbf{k}) \) are just complex numbers as magnetic vector is scalar particularly, \(\mathcal{U}_{\mu}(\textbf{k}) \in U(1) \cong \mathbb{C}^{1}\).
However, in our calculations, each \( \mathcal{U}_{\mu}(\textbf{k}) \) serves as a link variable and, collectively, forms a unitary matrix, \( \mathcal{U}_{\mu}(\textbf{k}) \in U(N) \cong \mathbb{C}^{N \times N} \), which acts on an \( N \)-dimensional degenerate subspace, extending the Abelian phase into a non-commuting gauge transformation.
}
\begin{figure}[t]
\includegraphics[width=1\linewidth, height=3.8 cm]{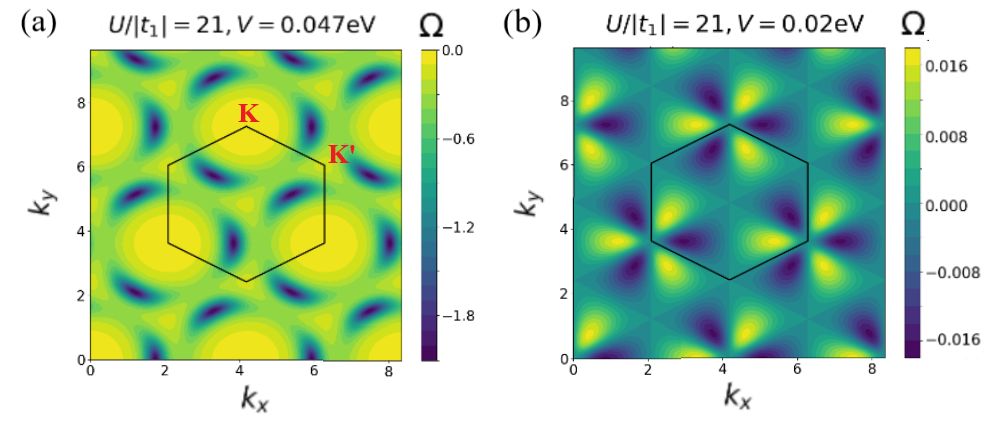}
\caption{Berry curvature maps calculated by using Eq. (\ref{chern_plaq}) for two selected values of displacement field $V=0.047$eV (a) and $V=0.02$eV (b). By integrating out the Berry curvature over the Brillouin zone one obtains $|C|=1$ (QAHI) for (a) and $|C|=0$ (AFI) for (b).}
\label{curvature}
\end{figure}
In this numerical calculation, we broke the entire Brillouin zone into small rectangular plaquettes and calculated the winding number for each of them which is given by,
\begin{align}
\delta C= &\frac{1}{2\pi i}Tr\big(\bm{\mathcal{A}}(1,2).{d\mathbf{k}} + \bm{\mathcal{A}}(2,3).{d\mathbf{k}} \nonumber\\
& + \bm{\mathcal{A}}(3,4).{d\mathbf{k}} + \bm{\mathcal{A}}(4,1).{d\mathbf{k}} \big).
\end{align}
{
Here the edges \(1\) to \(4\) define the loop around the plaquette in a counterclockwise direction typically defined on a Monkhorst-Pack grid.
Consequently, the above equation can be reformulated using link variables defined from overlaps of neighboring Bloch states. In the lattice formulation, the gauge potential \(\bm{\mathcal{A}} \cdot d\mathbf{k}\) is discretized as \(\ln \mathcal{U}\).}
 By carefully evaluating each \(\mathbf{k}\)-point across the entire Brillouin zone, we arrive at the final result presented below, 
\begin{equation}
 C^{\mathit{(2)}}_{n}=  \frac{1}{2\pi i} \oint{Tr (\bm{\mathcal{A}^{n}}.d\mathbf{k}})
 = \frac{1}{2\pi i} \sum_{\mathbf{k}}ln\hspace{0.05cm}  \bigg| \prod^{4}_{\mathcal{K}=1}\mathcal{U}(\mathcal{K})\bigg| ,
\end{equation}
where, \(|A| = det (A)\). We can write the compact form of the above term by summing up not all the \(\mathbf{k}\) points but all the small plaquettes,

\begin{equation}
 C^{\mathit{(2)}}_{n}= \frac{1}{2\pi i} \sum_{\substack{\square}} \ln\hspace{0.05cm}  \mathcal{U}_{\substack{\square}} ,
 \label{chern_plaq}
\end{equation}
where, \(\mathcal{U}_{\substack{\square}}\) represents the determinant of product of the four link variables.
\begin{figure}[b]
\includegraphics[width=1\linewidth, height=3.80cm]{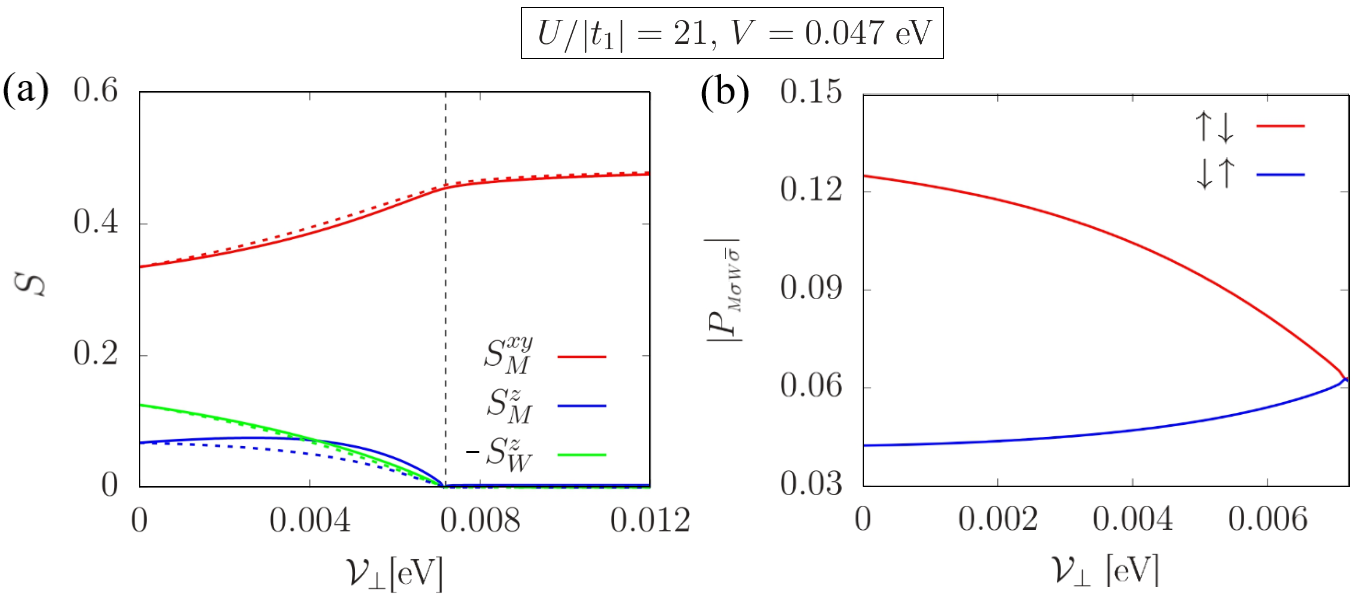}
\caption{ {(a) Magnetization components (in-plane 120$^\circ$ AF in MoTe\(_2\) and out-of-plane in both layers) as functions of the interlayer intersite Coulomb repulsion \(\mathcal{V}_\perp\) with (solid lines) and without (dashed lines) inclusion of interlayer coherence term \(\bigl(c_{i\ell\sigma}^\dagger c^{}_{j\ell'\sigma'}\bigr)\).(b)  Difference between the MoTe$_2$ and WSe$_2$
onsite energies with (solid lines) and without (dashed lines) inclusion of interlayer coherence term. The vertical dashed line indicates the transition from a topological phase to a trivial state.}
}
\label{updated_inter}
\end{figure}
Using Eq.~(\ref{chern_plaq}), the non-Abelian Berry curvature  \(\Omega(k_{x},k_{y})\) in \(\mathbf{k}\)-space is identified as the phase,  divided by the area of each small plaquette (\({\substack{\square}}\)). 
{ Taking into account the properties of the link-variable matrices, Eq. (\ref{curv_1}) can be reformulated as follows
\begin{equation}
\Omega_{\mu \nu} (\textbf{k}) = \ln \left( 
\frac{
\mathcal{U}_{\mu}(\textbf{k}) \mathcal{U}_{\nu}(\textbf{k} + \hat{\mu}) 
}{
\mathcal{U}_{\nu}(\textbf{k}) \mathcal{U}_{\mu}(\textbf{k} + \hat{\nu})
}
\right),
\label{curv_2}
\end{equation}
where, $\hat{\mu}$ is label for direction in momentum space and \(\boldsymbol{\mu \cdot \nu}\) represents area of \({\substack{\square}}\).
%{
In Fig. \ref{curvature} we show the calculated Berry curvature maps in momentum space for two selected values of the displacement field $V=0.047$eV and $V=0.02$eV, corresponding to the $|C|=1$ (QAHI) and $|C|=0$ (AFI), respectively.} 
When \(U\) is on the higher side, the phase diagram for the top three  moir\'{e} bands can be obtained by carefully comparing the results.

%%%%%%%%%%%%%%%%%%%%%%%%%%%%%%%%%%%%%%%%%%%%%%%%%%%%%%%%%%%%%%%%

\section{Influence of the interlayer coherence term}
{
Both formation and evolution of non-trivial topology in the canted antiferromagnetic state with gradual increase of $\mathcal{V}_{\perp}$ within the $t-U-\mathcal{V}$ model has been analyzed in Section \ref{sec:level3}. Here, we revisit the in-plane and out-of-plane magnetization for both layers, as presented in Fig. \ref{inter}(a), examining the impact of the \textit{interlayer coherence terms} originating from the mean-field decomposition of the interlayer Coulomb repulsion [cf. Eq. (\ref{ham_V_HF2})]. For comparison, in Fig. \ref{updated_inter} we provide the results obtained with (solid line) and without (dashed line) the inclusion of the additional terms.
As one can see in both situations the critical value of $\mathcal{V}_{\perp}$ at which the phase transition occurs remains unchanged, and the overall behavior is qualitatively similar. The values of the spin-flip interlayer mean-field parameters are provided in Fig. \ref{updated_inter}(b) within the topological region ($\mathcal{V}_{\perp}\lesssim 7$ meV). Close to the transition point ($\mathcal{V}_{\perp}\approx 7$ meV) the contribution resulting from the interlayer term to the effective Hamiltonian given by Eq. (\ref{ham_V_HF2}) is $\mathcal{V}_{\perp} |P_{MW\uparrow\downarrow}|\approx0.4$ meV, which is one order of magnitude smaller than the interlayer hopping that appears in the single-particle Hamiltonian ($t_{\perp}=4$ meV). This is the reason why the transition point to the topological phase is not much affected by the inclusion of the interlayer terms. The $\uparrow\downarrow$ and $\downarrow\uparrow$ terms have different values due the non-zero out-of-plane magnetizations in both layers.}

\bibliography{refs}% Produces the bibliography via BibTeX.

\end{document}